\documentstyle[aps,prl,psfig,floats,latexsym]{revtex}

\newcommand{\be}{\begin{equation}} \newcommand{\ee}{\end{equation}}
\def\(#1){(\ref{#1})} \newcommand{\bea}{\begin{eqnarray}}
  \newcommand{\eea}{\end{eqnarray}} 

\newcommand{\lav}{\left\langle} \newcommand{\rav}{\right\rangle}
 \newcommand{\ie}{{\it i.e.\,\,}}

 \newcommand{\tw}{t_{\rm w}}
 \newcommand{\twlim}{t_{\rm
    w}\to\infty}

\newcommand{\ts}{t_{\rm s}}

\newcommand{\teff}{T_{\rm eff}} 
\newcommand{\gdot}{\dot{\gamma}} 
\newcommand{\etal}{{\it et al.}}
\newcommand{\eps}{\epsilon}
\newcommand{\lae}{\stackrel{<}{\scriptstyle\sim}}
\newcommand{\dP}{\tilde{P}}
\newcommand{\cP}{\tilde{P}_{\rm age}}
\newcommand{\p}{P}

\newcommand{\ddtau}{\partial_\ta}
\newcommand{\ta}{\tau}
 
\begin{document}

\draft 

\twocolumn[\hsize\textwidth\columnwidth\hsize\csname @twocolumnfalse\endcsname

\title{Interrupted coarsening in a driven kinetically constrained
  Ising chain}

\author{Suzanne M. Fielding$^1$~\cite{present_address}}

\address{
Department of Physics and Astronomy, University of Edinburgh, Mayfield Road, Edinburgh, EH9 3JZ, United Kingdom\\
}

\maketitle

\begin{abstract}
  
  We introduce a driven version of the 1D kinetically constrained spin
  chain \cite{JaecEis91}. In its original undriven version, this model
  shows anomalous coarsening following a quench to a low temperature,
  with an equilibration time that diverges as $\sim\exp(1/T^2)$ for
  $T\to 0$.  We show that driving of constant rate $\gdot$ interrupts
  coarsening and stabilises the chain in a state analogous to that of
  a coarsening chain of age $1/\gdot$. We present an analytical theory
  for this steady state, and demonstrate it to be in excellent
  agreement with our simulation results.

\end{abstract}


]

\section{Introduction}
\label{sec:intro}

Glassy systems relax very slowly at low temperatures. They therefore
remain out of equilibrium for long times and exhibit ageing
\cite{BouCugKurMez98}: the time-scale for response to an external
perturbation (or for the decay of correlations) increases with the
``waiting time'' $\tw$ since the system was quenched to the low
temperature, and thus eventually far exceeds the experimental
time-scale. Time translational invariance is lost.

As a result of this dynamical sluggishness, glassy systems are highly
susceptible to steady external driving, even when the driving rate
$\gdot$ is small. (One example of $\gdot$ is shear rate in a
rheological system.) Typically, such driving interrupts ageing and
restores a time-translationally invariant (steady) non-equilibrium
state in which the time-scale defined by the inverse driving rate
plays a role analogous to the waiting time $\tw$ of the ageing
regime~\cite{CugKurLeDouPel97}.  This scenario was first investigated
in the context of neural networks\cite{HerGriSol86,CriSom87} and has
subsequently been reproduced in the diffusion of a particle in a
random potential\cite{Horner96,Thalmann98}; in driven mode-coupling
equations of the mean field p-spin model\cite{BerBarKur00}; and in a
driven version of Bouchaud's trap
model\cite{MonBou96,long_El,SolLeqHebCat97}.  The results of the
latter two studies were separately exploited to propose a general
framework for the study of ``soft glassy
materials''\cite{DerAjdDucLeq00,CloBorLei00,CipManBalWei00,KnaBelMunViaLeqHar00,RamCip01}
in which intrinsic rheological ageing is interrupted by driving (shear
straining) and loading (shear stress)\cite{CloBorLei00}.

In this paper, we introduce a driven version of another glassy model:
the 1D Ising chain with an asymmetric kinetic constraint. The original
undriven model (introduced by J\"{a}ckle and Eisinger\cite{JaecEis91})
shows anomalous slow coarsening (``ageing'') following a quench to a
low temperature $T\ll 1$, with an ergodic time that diverges as
$\tau\sim\exp(\rm{const}/T^2)$ for $T\to 0$, as solved exactly by
Sollich and Evans\cite{SolEva99}.  Particularly attractive features of
the model are i) that its glassiness emerges as a direct result of the
dynamical constraint (without the need for any underlying assumption
of quenched disorder) and ii) that it contains explicit spatial
interactions while being simple enough to allow analytical progress.
In what follows, our central result will be that steady driving
interrupts coarsening and stabilises the chain at an apparent age
$O(1/\gdot)$, consistently with the scenario described above.

Ageing and driven glassy systems in general violate the equilibrium
fluctuation dissipation theorem (FDT)\cite{Reichl80}.  Remarkably,
however, in the ageing limit $\twlim$ of many mean-field
\cite{CugKur94,CugKur93} and some non-mean-field
\cite{KobBar00,Barrat98,MarParRicRui98,RicParStaAre01} models, a
non-trivial modified FDT emerges and defines a non-equilibrium
temperature $\teff$\cite{CugKurPel97}.  Cugliandolo
\etal\cite{CugKurPel97} proposed that an equivalent temperature should
apply in the driven limit $\gdot \to 0$, and that $\teff(\gdot\to 0)$
and $\teff(\twlim)$ should coincide.  To date, however, the
evidence~\cite{BerBarKur00,BarBer01} for this is rather limited. An
added motivation for the present paper, therefore, is that the model
defined here can be used (in future work~\cite{upcoming_fdt}) to test
this scenario further via comparison of its FDT predictions with those
of the undriven model~\cite{CriRitRocSel00}.

The paper is structured as follows. In section~\ref{sec:undriven} we
summarise the results of\cite{SolEva99} for the coarsening dynamics of
the undriven chain. We then define the novel driving rules in
section~\ref{sec:driving}. We present simulation and analytical
results for the steadily driven chain in sections~\ref{sec:sim}
and~\ref{sec:theory} respectively, demonstrating that driving of rate
$\gdot$ halts coarsening at an effective age $1/\gdot$.  We summarise
and give an outlook for future work in section~\ref{sec:summary}.

\section{Coarsening in the undriven model}
\label{sec:undriven}

The model consists of a chain of $L$ Ising spins $s_i\in {0,1}$
($1<i\le L$) in a uniform field of unit magnitude, which is oriented
such that a spin's energy $E_i(s_i)=s_i$. Periodic boundary conditions
apply: the left neighbor of $s_1$ is $s_L$.  The dynamics are subject
to the following constraint: at any time, only those spins whose left
neighbor is up are allowed to flip.  For these ``mobile'' spins, the
rate of down-flips is $1$ while the rate of up-flips is
$\epsilon=\exp(-1/T)$. The {\em equilibrium} distribution is
unaffected by this constraint: detailed balance is obeyed and the
static distribution is the trivial one prescribed by the Hamiltonian
$H=\sum_{i=1}^L s_i$. In contrast, the dynamics are rendered very slow
at low temperatures, for which the equilibrium concentration of up
spins (that facilitate the dynamics) $c=\epsilon/(1+\epsilon)$ is
small.

In this section we review\cite{SolEva99}, which solved exactly the
relaxation of a chain prepared at a low temperature $T\ll 1$ via rapid
quench at time $\tw=0$ from a high initial temperature $T_i=\infty$.
At any time $\tw$ during the relaxation, the authors of\cite{SolEva99}
described the system's state using the concept of ``domains''.  As
shown by the vertical lines in
\begin{center}
$...1|\cdot\cdot1|\cdot\cdot\cdot\cdot\cdot\cdot\cdot\cdot1|\cdot1|\cdot\cdot\cdot\cdot1|\cdot\cdot\cdot1|1|1|\cdot1...$
\end{center}
(in which the 0 spins are represented by ``$\cdot$'' for clarity) a
domain is defined as an up-spin, and all the down-spins separating it
from the next up-spin to the left.

Immediately after the quench, the average domain length $\bar{d}\equiv
1/c=O(1)$.  The system then relaxes towards the low temperature's
equilibrium state in which $\bar{d}=O(1/\epsilon)$, and thus in which,
in the limit $\eps\to 0$, there is zero probability of finding an up
spin in any finite length of chain. Hence, in this limit ($\epsilon
\to 0$ at fixed chain length) the down-flipping of spins is
irreversible, and the relaxation comprises a coarsening process in
which adjacent domains progressively coalesce with one another.

The mechanism for this coalescence is as follows. Consider, at some
time after quench, a domain of length $d\ll 1/\epsilon$ together with
its left-bounding up-spin.  For the purposes of the present argument,
we assume that the left-bounding spin is ``clamped'' and consider how
the right-bounding up-spin relaxes. Because of the constraint, before
the relaxation can occur a facilitating up-spin has to be generated
immediately to the left of this spin, via a propagation of the up
state rightwards from the left-bounding up-spin.  The relaxation is
thus impeded by an energy barrier, the height of which is the maximal
number of spins that are ever up within the (original) domain at any
instant during this relaxation process. A central result
of\cite{SolEva99} is that for domain lengths $2^{n-1}<d\le 2^n$ this
barrier scales as $n$, leading to a relaxation time-scale
$O(\epsilon^{-n})$.

Hence in the limit $\eps\to 0$ the dynamics comprise well separated
stages, the $n^{\rm th}$ of which has time-scale $\epsilon^{-n}$ and
results in domains of index $n$ being destroyed by coalescence with
their right neighbors.  In logarithmic time, $\nu_{\rm w}= -\log
\tw/\log\eps=T\log(\tw)$, the $n^{\rm th}$ stage collapses to the
point $\nu_{\rm w}=n$.  The average domain size $\bar{d}$ thus
exhibits step-wise increases at successive integer values of $\nu_{\rm
  w}$, as seen in figure 1 of\cite{SolEva99}.

Within this coarsening regime ($\bar{d}\ll 1/\epsilon$), the full
domain length distribution $P(d)$ can be calculated using an exact
independent interval treatment~\cite{BraDerGod94} which states that no
correlations can build up in the length of adjacent domains provided
none are present in the initial state. At stage $n$ the distribution
obeys~\cite{SolEva99}
\be
\ddtau \p(d,\ta) = \sum_{ 2^{n-1}< d'\leq 2^n }
\p(d-d',\ta)\,[-\ddtau \p(d',\ta)]\; 
\label{eqn_motion}
\ee
in which the rescaled time $\ta=t\eps^n$ can take any positive value
$\tau>0$ in the limit $\eps\!\to\! 0$. Equation~\ref{eqn_motion}
describes the coalescence of the ``active'' domains of length $d'\leq
2^n$ with neighboring domains of length $d-d'$. Its initial condition
is the domain length distribution at the end of stage $n-1$ of the
dynamics, denoted $\p_n(d)$ = $\p(d,\ta\!\to\!  0)$. Using generating
functions it can be shown that
\bea
\label{eqn:series}
P_{n+1}(d)&=&P_n(d)+\sum_{d'=1}^{d-1} P_n(d-d')P_n^{\rm\, act}(d')\nonumber\\
          & &-\frac{1}{2}\sum_{d'=1}^{d-1}P_n^{\rm\, act}(d-d')P_n^{\rm\, act}(d')+...
\eea
in which $P_n^{\rm\, act}(d)$ is the active part of the distribution
(zero for $d>2^n$) and in which... denotes a series of convolutions of
increasing order.  Equation~\ref{eqn:series} holds only for $d>2^n$:
all active domains ($d\le 2^n$) disappear in the $n^{\rm th}$ stage.

The weight of the distribution shifts to larger $d$ at each stage of
coarsening. See figure 2 of \cite{SolEva99}.  A scaling limit is
approached for large stage number $n$: the re-scaled distribution
$\tilde{P}_n(x=d/2^{n-1})=2^{n-1}P_n(d)$ converges to the limit
$\cP(x)$ which obeys, for $x>2$, the scaling counterpart of
equation~\ref{eqn:series}:
\bea
\label{eqn:series_scaling}
\frac{1}{2}\cP\left(\frac{x}{2}\right)&=&\cP(x)+\int_{0}^{x}dx' \cP(x-x')\cP^{\rm\, act}(x')\nonumber\\
          & &-\frac{1}{2}\int_{0}^{x}dx'\cP^{\rm\, act}(x-x')\cP^{\rm\, act}(x')+...
\eea
In~\cite{SolEva99} the exact solution for $\cP(x)$ was shown (via the
re-summed Laplace transform of~\ref{eqn:series_scaling}) to be
\be
\label{eqn:dom_length_distr_scaling}
\cP(x)=\sum_{m=1}^{\infty}\frac{(-1)^{m-1}}{m!}\int_1^\infty \prod_{r=1}^m \frac{dx_r}{x_r}\delta\left(\sum_{s=1}^{m}x_s-x\right).
\end{equation}

\section{Definition of the driven model}
\label{sec:driving}

In this section we incorporate non-Hamiltonian driving into the model.
As a preliminary step, though, we redefine the {\em relaxational}
dynamics slightly, extending state space such that each spin
$s_i\in{-1,0,1}$ and re-defining the Hamiltonian $H=\sum_{i=1}^L
|s_i|$. The uniform field has thus been replaced by a potential well
for each spin, with a minimum at $s_i=0$. As before the dynamics are
constrained: only those spins for which the left neighbor has a value
$1$ or $-1$ are allowed to flip via the usual thermal processes in
which the transition rate for $s_i:1\to 0$ and for $s_i:-1\to 0$ is 1,
and for $s_i:0\to 1$ and $s_i:0\to -1$ is $\epsilon$.

So far, of course, the model can be exactly mapped onto the original
one by a trivial relabelling $s_i=-1\to s_i=1$ and re-scaling $\eps \to
\eps/2$. Our motivation for introducing the $-1$ state is to make a
loose analogy with glassy rheological
models\cite{SolLeqHebCat97,DerAjdLeq99,HebLeq98} in which a local
state of high energy (here $|s_i|=1$) can have either positive or
negative local stress (here $|s_i|=+1$, $|s_i|=-1$). If we define a
global stress $\sigma=\frac{1}{L}\sum_{i=1}^L s_i$, the $-1$ state
allows a state of {\em macroscopically} zero stress, which still has
internal local stresses (some positive, some negative) and a non-zero
rate of internal dynamical rearrangements.

We now incorporate steady driving into this three state version.
Loosely this mimics, in a stochastic way, the standard rheological
experiment of applying shear strain of constant rate $\gdot$. To do
this, we impose a flip rate of $\gdot$ for $s_i:-1\to 0$ and $\gdot$
for $s_i: 0\to 1$.  This is {\em additional} to the constrained rates
defined above and {\em free of the kinetic constraint}.  The driving
rates $\omega$ for this extended model can therefore be summarised as
follows:
\bea
\omega(s_i:0\to 1)&=& |s_{i-1}|\eps +\gdot\nonumber\\
\omega(s_i:0\to -1)&=& |s_{i-1}|\eps\nonumber\\
\omega(s_i:1\to 0)&=&  |s_{i-1}|\nonumber\\
\omega(s_i:-1\to 0)&=&  |s_{i-1}|+\gdot,\nonumber\\
\eea
in which periodic boundary conditions impose $s_{0}=s_{L}$.

This stochastic straining clearly tends to increase the global stress,
as required intuitively. We note, though, that our stochastic rules
only make sense for $\gdot\ge 0$. For negative $\gdot$ we would
redefine the driven contribution to the rates as equal to $|\gdot|$
for the transition $s_i:1\to 0$ and $|\gdot|$ for $s_i:0\to -1$.  The
model is in this sense singular at $\gdot=0$.

\section{Simulation results for the steadily sheared model}
\label{sec:sim}

We simulated the driven chain using a waiting time Monte Carlo
technique combined with a binary search algorithm for locating the
mobile spins, following \cite{SolEva99}.  For each run we initialised
the chain either in equilibrium (with $\gdot=0$) at a low temperature
$T=-1/\log\epsilon\ll 1$, or by quenching to $T$ from $T=\infty$. For
the quenched case, we then let the system relax according to the
undriven rules ($\gdot=0$) until a start-up time $\ts$, when we set
$\gdot$ to the non-zero, constant value of interest. Equilibrium
initialisation formally corresponds to a quenched chain subsequently
allowed to relax until $\ts=\infty$, and in this case we applied the
non-zero $\gdot$ from the start of the simulation. In order to explore
the hypothesis that driving restores a steady state analogous to the
state of a coarsening chain of age $\tw=1/\gdot$, we chose values of
$1/\gdot$ corresponding to the waiting times studied in
\cite{SolEva99} for the undriven chain.  Specifically, therefore, we
are interested in the low temperature limit in which $\eps\to 0$,
$\gdot\to 0$, at fixed values of
$\nu=\log(\gdot)/\log(\eps)=T\log(1/\gdot)$ that are large compared to
$1$ (weak driving), but small enough that the system remains far from
equilibrium. (To avoid possible confusion we note that $\nu$ becomes
large as $\gdot$ becomes {\em small}, since $\gdot=\eps^{\nu}$ with
$\eps\ll 1$. Although this is at first sight a counter intuitive way
to characterise the shear rate, we chose this particular definition
for $\nu$ as the closest possible analogy that of $\nu_{\rm w}$ for
the undriven chain.)

\begin{figure}[h]
\centerline{\psfig{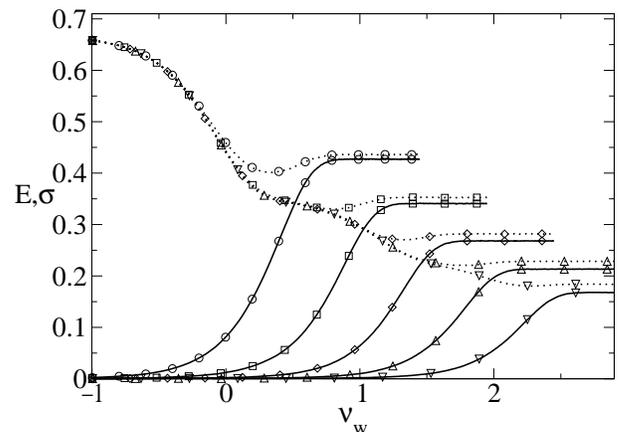}}
\caption{Stress $\sigma$ (solid  lines) and energy $E$ (dashed lines) {\it vs.} scaled time $\nu_{\rm w}=-\log(\tw)/\log(\eps)$ following a quench at time $\tw=0$, with driving commenced also at $\ts=0$; for all curves $\eps=0.01$. The parameter $\nu\equiv \log\gdot/\log\eps$ has values $0.5$ ($\bigcirc$), $1.0$ ($\Box$) $1.5$ ($\Diamond$) $2.0$ ($\bigtriangleup$) and $2.5$ ($\bigtriangledown$). Each curve was obtained from a single run for a chain of length  $L=2^{16}$.
\label{fig:time_ev} } 
\end{figure}

In each run we monitored the stress $\sigma$ and total energy $E$ as
functions of time $\tw$. Results for the quenched initial condition
with start-up time $\ts=0$ are shown in figure~\ref{fig:time_ev}, for
$\eps=0.01$ and various values of $\nu$. As $\nu$ becomes larger
($\gdot$ smaller) we can make the following observations. At early
times, $E$ and $\sigma$ have time evolutions that are independent of
$\gdot$, and that (we have checked) are the same as those of an
undriven chain.  In contrast, after a crossover at time $O(1/\gdot)$,
$E$ and $\sigma$ approach steady-state values.  We have checked by
repeating the simulation for the different initial conditions
described above that these steady values do not depend upon the
initial state or (for the quenched case) the start-up time $\ts$. For
the remainder of the paper, we shall be concerned only with the
ultimate steady state, and not the kinetics of its formation.

\begin{figure}[h]
\centerline{\psfig{figure=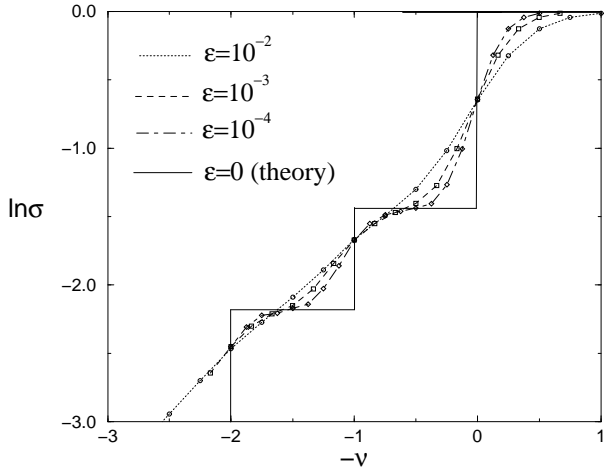,width=8cm}}
\caption{Steady state flow curves  plotted on a log scale vs. $-\nu=-\log\gdot/\log\epsilon$. Simulation results for three values of $\epsilon=\exp(1/T)$ are shown, obtained from a single run for a spin chain of length $L=2^{16}$. Bold line: theoretical prediction for $T\to 0$.
\label{fig:driven_flow_curves} } 
\end{figure}

The steady state stress is re-plotted in
figure~\ref{fig:driven_flow_curves} as a function of $-\nu$ (which
increases with $\gdot$) for various small values of $\epsilon\ll 1$.
(In rheological parlance, $\sigma(\gdot)$ is the flow curve.) It
appears to be approaching a step-like function as temperature is
tracked towards zero, comprising plateaux separated by jump-wise
discontinuities at integer values of $\nu$.

For all the steady states studied, we found the concentration of $-1$
spins to be small ($O(\epsilon)$), and hence that (to within such
corrections) the stress, the energy, and the concentration of $+1$
spins coincide (consistently with the results of
figure~\ref{fig:time_ev}).  Using the usual domain description (in
which for definiteness we neglect the $-1$ spins, taking only $1$
spins to constitute the domain boundaries), we have then
$\sigma=1/\bar{d}$ where $\bar{d}$ is the average domain length.

We note the striking similarity between the dependence of
$\log\bar{d}$ upon scaled waiting time $\nu_{\rm w}=T\log\tw$ in the
coarsening chain (figure 1 of\cite{SolEva99}), and the dependence of
$\log\sigma=-\log\bar{d}$ upon the scaled driving rate
$-\nu=-T\log(1/\gdot)$ in the steadily driven chain
(figure~\ref{fig:driven_flow_curves} of this paper). This already
gives us a strong indication that the steady state of a chain driven
at rate $\gdot$ is analogous to the state of a coarsening chain of age
$\tw=1/\gdot$.

\begin{figure}[h]
\centerline{\psfig{figure=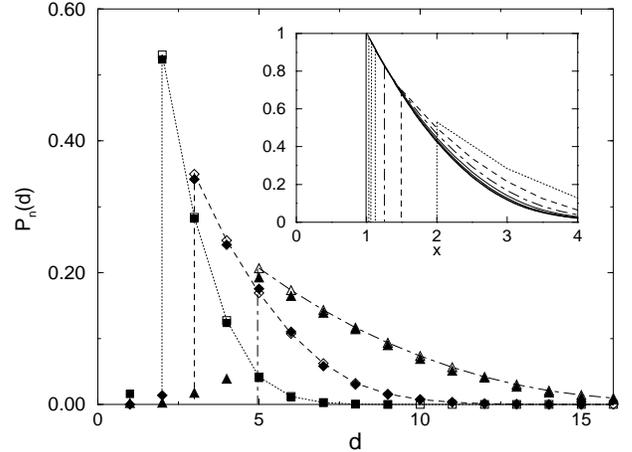,width=8cm}}
\caption{Domain length distributions $P_n(d)$ on the plateau $\nu=n-1/2$ of the flow-curves. Open symbols and lines:  theoretical results for $n=1$ (squares), 2 (diamonds), 3 (triangles). Full symbols: simulation results for a chain of length $L=2^{16}$ and $\epsilon=10^{-4}$ ($n=1,2$) and $\epsilon=10^{-3}$ ($n=3$). Inset: scaled predictions $2^{n-1}P_n(d=2^{n-1}x)$ vs. $x$ for $n=1,...,16$.  
\label{fig:driven_domain_distr} } 
\end{figure}

This hypothesis is confirmed by our simulation data for the full
domain length distribution $P(d)$ on the developing plateaux of the
flow curves.  In particular, we find that the distribution for the
steadily driven chain at a given $\nu$ is closely analogous to that of
a coarsening chain for $\nu_{\rm w}=\nu$. For a given $\nu$ (or
$\nu_{\rm w}$), both display a discontinuity at the same $\nu$
($\nu_{\rm w}$) dependent cutoff and have very similar averages.  Both
shift abruptly to larger values of $d$ as $\nu$ ($\nu_{\rm w}$)
crosses successive integers, but are unchanged as $\nu$ ($\nu_{\rm
  w}$) is swept between integers. In order to maintain the closest
notational analogy with\cite{SolEva99}, we denote by $P_n(d)$ the
distribution $P(d)$ in the limit of small $\eps$ for values of $\nu$
such that $n-1<\nu<n$.  Our results for $P_n(d)$ for
$\nu=0.5,\,1.5,\,2.5$, corresponding to $n=1,\,2,\,3$, are shown in
figure~\ref{fig:driven_domain_distr} and are, as just noted, very
similar to the counterpart results of figure 2 of\cite{SolEva99} for
the coarsening chain.

In the coarsening chain, the discontinuous shift of $P(d)$ as
$\nu_{\rm w}$ crosses successive integers arose from the waiting time
crossing the time-scales for successive coarsening stages. Likewise in
the driven chain it arises from the inverse driving rate crossing
these same time-scales. (See the theory section~\ref{sec:theory} for
more details.)

These simulation results therefore show that (consistently with the
phenomenology of other driven glassy models) driving interrupts
coarsening and stabilises the chain in a state that is strikingly
analogous to that of a coarsening chain of apparent age $O(1/\gdot)$.

\section{Theory}
\label{sec:theory}

We now present a theory aimed at calculating the steady-state
distribution $P_n(d)$ just defined.  We assume from the outset that no
correlations exist between the lengths of adjacent domains. We also
assume (consistently with our above remarks) that the concentration of
$-1$ spins is negligible. We will return to justify these assumptions
in more detail at the end of this section.

Consider, then, a chain with an $O(1)$ population of $+1$ spins and a
complementary $O(1)$ population of $0$ spins. In the absence of
driving, the only process affecting the chain would be the coarsening
described above, with down-flipping of $1$ spins occurring irreversibly
(for fixed chain length and $\eps\to 0$), and the average domain
length increasing as domains of length $2^{n-1}<d\le 2^n$ are
destroyed on a time-scale $O(\epsilon^{-n})$ by coalescence with their
right neighbors.

In the steadily driven chain, this coarsening process is balanced by
intra-domain driven up-flipping of spins $s_i:0\to 1$. For the finite
domain lengths to which we shall restrict ourselves, this driven
domain intersection occurs on a time-scale $O(1/\gdot)\equiv
O(\epsilon^{-\nu})$.  For a non-integer value of $\nu$ such that
$n-1<\nu<n$, this time-scale sits between the time-scales
$\epsilon^{-(n-1)}$ and $\epsilon^{-n}$ for the adjacent coarsening
stages $n-1$ and $n$ (recall section~\ref{sec:undriven}), and
separates from them in the limit $\epsilon\to 0$.  

On this driving time-scale, therefore, any domain of index $n'\le n-1$
(present either as a remnant of the initial condition, or as a result
of a driven up-flip $0\to 1$ within an existing domain of length
$d>2^{n-1}$ a distance $d\le 2^{n-1}$ from the domain boundary) must
relax infinitely quickly. We thus expect $\lim_{\eps\to 0} P_n(d)=0$
for such domain lengths $d\le 2^{n-1}\equiv d_c$. This defines the
cut-off length $d_c$ observed in the simulation data above.

Domains labelled by $n'>n-1$, \ie of length $d>d_c$, on the other
hand, coarsen infinitely slowly on the driving time-scale: the only
dynamical processes that can affect these ``long'' domains are those
initiated by driven up-flip of an intra-domain 0 spin. This up-flip
can occur at a distance that is either $\le d_c$ or $>d_c$ from either
end, giving 4 separate cases:

\begin{itemize}

\item
  
  {\bf A.} If the original long domain was longer than $2d_c+1$, and
  if the up-flip occurred at a distance of at least $d_c+1$ from both
  ends, we see creation of two shorter domains that are still both
  ``long'' in the sense that $d>d_c$:
\begin{center}
\vspace{-0mm}
$...1\cdot\cdot\,\cdot\,\cdot\cdot\cdot\cdot\cdot\cdot\cdot\cdot\cdot\cdot\cdot\cdot\cdot\cdot\cdot\cdot\cdot\cdot\cdot\cdot\cdot\cdot\cdot\cdot\cdot1\cdot\cdot\cdot\cdot\cdot\cdot\cdot\cdot\cdot\cdot\cdot\cdot\cdot\cdot\cdot\cdot\cdot\cdot1...$
\vspace{-2mm}

$\downarrow$

\vspace{-2mm}
$...1\cdot\cdot\cdot\cdot\cdot\cdot\cdot\cdot\cdot\cdot\cdot\cdot\cdot\cdot\cdot\cdot1\cdot\cdot\cdot\cdot\cdot\cdot\cdot\cdot\cdot\cdot\cdot1\cdot\cdot\cdot\cdot\cdot\cdot\cdot\cdot\cdot\cdot\cdot\cdot\cdot\cdot\cdot\cdot\cdot\cdot1...$
\vspace{-0mm}

\end{center}
(For definiteness all the diagrams in this section assume an arbitrary
cutoff value $d_c=4$.)

Considering the chain as a whole, such processes lead to the
destruction of domains of length $d>2d_c+1$ at a rate
\be
\label{eqn:rate1}
\gdot(d-2d_c-1)P_n(d)
\end{equation}
and to the creation of domains of length $d>d_c$ at a rate
\be
\label{eqn:rate2}
2\gdot\sum_{d'=d+d_c+1}^\infty P_n(d').
\end{equation}

\item

  {\bf B.} If the up-flip occurred within a distance $d_c$ of the
  left-hand end of the original domain, but at a distance greater than
  $d_c+1$ from the right hand end we see a process such as

\begin{center}
\vspace{-0mm}
$...1\cdot\cdot\cdot\cdot\cdot\cdot\cdot\cdot\cdot\cdot\cdot\cdot\cdot\cdot\cdot\cdot\cdot\cdot\cdot\cdot\cdot\cdot\cdot\cdot\cdot\cdot\cdot\cdot\cdot1\cdot\cdot\cdot\cdot\cdot\cdot\cdot\cdot\cdot\cdot\cdot\cdot\cdot\cdot\cdot\cdot\cdot\cdot 1...$
\vspace{-2mm}

$\downarrow$

\vspace{-2mm}
$...1\cdot\cdot\cdot 1\cdot\cdot\cdot\cdot\cdot\cdot\cdot\cdot\cdot\cdot\cdot\cdot\cdot\cdot\cdot\cdot\cdot\cdot\cdot\cdot\cdot\cdot\cdot\cdot\cdot1\cdot\cdot\cdot\cdot\cdot\cdot\cdot\cdot\cdot\cdot\cdot\cdot\cdot\cdot\cdot\cdot\cdot\cdot 1...$
\vspace{-2mm}

$\downarrow$

\vspace{-2mm}
$...1\cdot\cdot\cdot\cdot\cdot\cdot\cdot\cdot\cdot\cdot\cdot\cdot\cdot\cdot\cdot\cdot\cdot\cdot\cdot\cdot\cdot\cdot\cdot\cdot\cdot\cdot\cdot\cdot\cdot1\cdot\cdot\cdot\cdot\cdot\cdot\cdot\cdot\cdot\cdot\cdot\cdot\cdot\cdot\cdot\cdot\cdot\cdot 1...$
\vspace{-0mm}

\end{center}
which does not need to be considered further since it results in no
net change.

\item
  
  {\bf C.} If the up-flip occurred within a distance $d_c$ of the
  right-hand end of the original domain, but at a distance greater
  than $d_c$ from the left hand end, the ``short'' (right) sub-domain
  will relax immediately by coalescence with its right neighbor:
\begin{center}
\vspace{-0mm}
$...1\cdot\cdot\cdot\cdot\cdot\cdot\cdot\cdot\cdot\cdot\cdot\cdot\cdot\cdot\cdot\cdot\cdot\cdot\cdot\cdot\cdot\cdot\cdot\cdot\cdot\cdot\cdot\cdot1\cdot\cdot\cdot\cdot\cdot\cdot\cdot\cdot\cdot\cdot\cdot\cdot\cdot\cdot\cdot\cdot\cdot\cdot1...$
\vspace{-2mm}

$\downarrow$

\vspace{-2mm}
$...1\cdot\cdot\cdot\cdot\cdot\cdot\cdot\cdot\cdot\cdot\cdot\cdot\cdot\cdot\cdot\cdot\cdot\cdot\cdot\cdot\cdot\cdot\cdot\cdot\cdot1\cdot\cdot1\cdot\cdot\cdot\cdot\cdot\cdot\cdot\cdot\cdot\cdot\cdot\cdot\cdot\cdot\cdot\cdot\cdot\cdot1...$
\vspace{-2mm}

$\downarrow$

\vspace{-2mm}
$...1\cdot\cdot\cdot\cdot\cdot\cdot\cdot\cdot\cdot\cdot\cdot\cdot\cdot\cdot\cdot\cdot\cdot\cdot\cdot\cdot\cdot\cdot\cdot\cdot\cdot1\cdot\cdot\cdot\cdot\cdot\cdot\cdot\cdot\cdot\cdot\cdot\cdot\cdot\cdot\cdot\cdot\cdot\cdot\cdot\cdot\cdot1...$
\vspace{-0mm}

\end{center}
leaving two ``long'' ($d>d_c$) domains, with the boundary between them
shifted to the left. 

Processes such as this lead to the creation of domains of length $d$
at a rate
\begin{equation}
\gdot \sum_{d'=d+1}^{d+d_c}P_n(d')+\gdot \sum_{d'=1}^{d_c}\left[\sum_{d''=d_c+d'+1}^{\infty}P_n(d'')\right]P_n(d-d')\nonumber
\end{equation}
and to the destruction of domains of length $d$ at a rate
\bea
\label{eqn:create}
&&\gdot\Theta(d-2d_c)d_cP_n(d)+\gdot\Theta(2d_c-d+1)(d-d_c-1)P_n(d)\nonumber\\
&&+\gdot P_n(d)\left[\sum_{d'=d_c+1}^{2d_c}(d'-d_c-1)P_n(d')+d_c\sum_{d'=2d_c+1}^{\infty}P_n(d')\right]\nonumber
\end{eqnarray}
where the discrete Theta function is defined by
\be
\begin{array}{lcll} 
\Theta(n-m)&=&1&\mbox{for $n>m$}\\
           &=&0&\mbox{for $n\le m$}.
\end{array}
\end{equation}

\item 
  
  {\bf D.} If the original domain was of length $d\le 2d_c$, both the
  sub-domains could be short ($d\le d_c$). For this class as a whole
  (\ie averaging over the position of the up-flipped spin) there are
  two sub-classes of possible outcome, which by symmetry each occurs
  with a probability one half. First, the freshly flipped up-spin
  could relax before the right-bounding spin of the original domain
  and we would see no net change.  Alternatively, we could see
\begin{center}
\vspace{-0mm}
$...1\cdot\cdot\cdot\cdot\cdot1\cdot\cdot\cdot\cdot\cdot\cdot\cdot\cdot\cdot\cdot\cdot\cdot\cdot\cdot\cdot\cdot\cdot\cdot 1...$
\vspace{-2mm}

$\downarrow$

\vspace{-2mm}
$...1\cdot\cdot1\cdot\cdot1\cdot\cdot\cdot\cdot\cdot\cdot\cdot\cdot\cdot\cdot\cdot\cdot\cdot\cdot\cdot\cdot\cdot\cdot 1...$
\vspace{-2mm}

$\downarrow$

\vspace{-2mm}
$...1\cdot\cdot1\cdot\cdot\cdot\cdot\cdot\cdot\cdot\cdot\cdot\cdot\cdot\cdot\cdot\cdot\cdot\cdot\cdot\cdot\cdot\cdot\cdot 1...$
\vspace{-2mm}

$\downarrow$

\vspace{-2mm}
$...1\cdot\cdot\cdot\cdot\cdot\cdot\cdot\cdot\cdot\cdot\cdot\cdot\cdot\cdot\cdot\cdot\cdot\cdot\cdot\cdot\cdot\cdot\cdot\cdot 1...$
\vspace{-0mm}

\end{center}
which is essentially ``aided coalescence''.  For the chain as a whole,
processes such as these lead to destruction of domains of length $d$
at a rate
\bea
\gdot P_n(d)\frac{1}{2}\left(2d_c-d+1\right)&&\nonumber\\
+\frac{1}{2}P_n(d)\gdot\sum_{d'=d_c+1}^{2d_c}P_n(d')\left(2d_c-d'+1\right)&&\nonumber
\end{eqnarray}
and creation at a rate
\be
\frac{\gdot}{2}\sum_{d'=d_c+1}^{2d_c}\left(2d_c-d'+1\right)P_n(d')P_n(d-d').\nonumber
\end{equation}

\end{itemize}
Combining all these processes we get an evolution equation $\partial_t
P_n(d)=...$ for the ``long'' domains $d>d_c$, which we set equal to
zero (steady state) and solve numerically using an iterative
procedure. The solutions for $n=1,\,2,\,3$ for which $d_c=1,\,2,\,4$
respectively are marked as open symbols
figure~\ref{fig:driven_domain_distr} and give excellent agreement with
the simulation results. We also used these solutions to calculate the
stress, $\sigma=1/\bar{d}$. As expected, this exhibits discontinuous
jumps at integer values of $\nu$ as the driving time-scale crosses
successive coarsening time-scales and $P(d)$ shifts discontinuously to
larger $d$. It is marked as the solid line in
figure~\ref{fig:driven_flow_curves} and again agrees excellently with
the simulation data.

\begin{figure}[h]
\centerline{\psfig{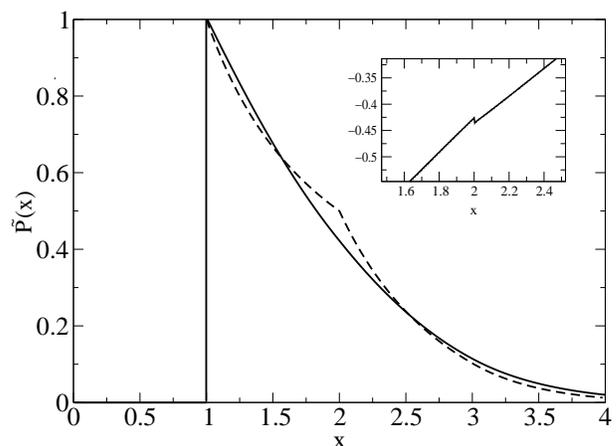}}
\caption{Scaling distribution $\tilde{P}(x)$ for the driven chain (solid line) and $\tilde{P}_{\rm age}(x)$ for the coarsening chain (dashed line). Inset: derivative $d\tilde{P}/dx$ for the driven distribution, showing a slight discontinuity (too small to be discernible in the undifferentiated data of the main figure) at $x=2$.
\label{fig:compare} } 
\end{figure}

As $d_c\to\infty$, a scaling limit $d_cP(d)=\dP (x=d/d_c)$ is
approached. See the inset of figure~\ref{fig:driven_domain_distr}; and
figure~\ref{fig:compare}. Taking the limit $d_c\to\infty$ at fixed
$x=d/d_c$ in the steady state equation just derived, we find that this
scaling state must obey the equation
\bea
\label{eqn:scaling_eqn}
0&=&-g(x)\dP(x)+2\int_x^\infty dx'\dP(x')-\int_x^{x+1}dx'\dP(x')\nonumber\\
 & &+\int_0^2 dx'f(x')\dP(x-x').
\eea
We note for use below that the function $f(x)$ (thus defined) is
discontinuous at $x=1$, and the first derivative $g^{(1)}(x)$ of $g$
is likewise discontinuous at $x=2$. In principle,
equation~\ref{eqn:scaling_eqn} contains all the information needed to
calculate $\dP$ analytically. For the counterpart state $\cP(x)$ in
the undriven chain, the closed expression
(equation~\ref{eqn:dom_length_distr_scaling}) was found
\cite{SolEva99} as the self-consistent solution of a simple algebraic
relation between the Laplace transforms $G$ and $H$ of $\cP(x)$ and
$\cP(x)\Theta(2-x)$ respectively.  The corresponding transform of
equation~\ref{eqn:scaling_eqn} for the driven case is a complicated
differential relation between $G$, $H$, $G^{(1)}$ and $H^{(1)}$, and
we have been unable find a self-consistent analytic solution.
%
%
However our numerical results (see figure~\ref{fig:compare})
demonstrate that $\dP(x)$ is (as expected) very similar to its
undriven counterpart $\cP(x)$: both have a unit Heaviside
discontinuity at $x=1$ and show similar decay for $x>1$. Although the
discontinuity in the first derivative $\cP^{(1)}(x)$ at $x=2$
(strongly apparent in the dashed curve of figure~\ref{fig:compare}) is
less noticeable in the solid curve for the driven state $\dP$, it is
revealed by numerical differentiation in the inset of
figure~\ref{fig:compare}. 

It was shown in\cite{SolEva99} that the ageing scaling distribution
$\cP$ has a finite discontinuity in its $k^{\rm th}$ derivative at
$x=k+1$ for all integer $k\ge 0$. We have already seen numerically
that the driven scaling distribution shares the discontinuities for
$k=0,1$.  We shall now outline an analytical argument which can be
used to show that in fact the driven distribution shares {\em all} of
these discontinuities. We confine ourselves to $k\ge 1$ since our
analysis has already captured the discontinuity in $\dP$ itself at
$x=1$.  Differentiating equation~\ref{eqn:scaling_eqn} once we get
\bea
\label{eqn:scaling_eqn_diff1}
-g(x)\dP^{(1)}(x)&=&g^{(1)}(x)\dP(x)+\dP(x)+\dP(1+x)\nonumber\\
            & &-\int_0^2 dx'f(x')\dP^{(1)}(x-x').
\eea
On the right hand side (RHS) of this expression, $g^{(1)}(x)$ is
discontinuous at $x=2$ (as noted above), while the integral over the
infinite discontinuity $\delta(x-x'-1)$ of the differential
$\dP^{(1)}(x-x')$ in the last term's integrand gives
$\Theta(3-x)f(x-1)$, which is discontinuous at $x=2$.  All other terms
on the RHS are continuous for all $x>1$.  Hence $P^{(1)}(x)$ has a
finite discontinuity at $x=2$. By performing successive
differentiations, we can extend this argument to arbitrarily high $k$:
$\dP^{(k)}(x)$ has a finite discontinuity at $x=k+1$ for all $k\ge 0$.

Our central result can therefore be summarised as follows. We have
shown that a steadily driven chain approaches a scaling state in the
limit of small $\gdot$. We have shown that this scaling state is
strikingly analogous to the counterpart scaling state of a coarsening
chain of age $\tw=1/\gdot$.  In particular, both states have a unit
Heaviside discontinuity at $x=1$, and (more generally) a finite
discontinuity in $\dP^{(k)}(x)$ at $x=k+1$.

Despite the striking qualitative similarities in the distributions
$\dP(x)$ and $\cP(x)$, there are obvious quantitative discrepancies.
This is not inconsistent with the phenomenology of other glassy
models: the ageing state of Bouchaud's trap model~\cite{MonBou96}, for
example, is analogous but not identical to its steadily driven
counterpart~\cite{SolLeqHebCat97}. It is nonetheless instructive to
consider the origin of the discrepancies.  An obvious candidate is our
introduction in the driven model of the $-1$ spin state.  However, in
steady state the population of such spins is small. Indeed, we have
checked that domain distributions produced by driven simulations
without the $-1$ state agree with those that include the $-1$ state to
within $O(\eps)$.

The discrepancies must therefore be of dynamical origin.  During
coarsening, the only process is domain coalescence.  Therefore
information can only propagate up the distribution (to larger $d$). In
this way the scaling limit $\cP(x)$ is completely determined by
$P(x')$ for $x'<x$, as seen in equation~\ref{eqn:series_scaling}. In
contrast, in the driven chain $\dP(x)$ is connected to all
$x'\to\infty$, since driven domain intersection acts on domains of all
lengths; see equation~\ref{eqn:scaling_eqn}. Furthermore, in the
ageing chain the domain lengths set the dynamical rates only via the
coarsening time-scales $\eps^{-n}$. In contrast, the driving dynamics
are further sensitive to domain length through the additional feature
that a longer domain is more likely to be intersected.

The most apparent difference between $\cP(x)$ and $\dP(x)$ is the much
smaller discontinuity in $\dP^{(1)}(x)$ at $x=2$.  This can be
explained as follows. The discontinuity in $\dP^{(1)}(x)$ arises from
an interplay between the discontinuity in $g^{(1)}(x)$ at $x=2$ and in
$f(x)$ at $x=1$ (recall equation~\ref{eqn:scaling_eqn_diff1}).
Physically, $g(x)$ is the rate at which domains of scaling length $x$
are destroyed. This is discontinuous because driven intersection of a
domain of scaled length $x<2$ can produce two active domains (scaling
length $<1$), whereas only one can be active for $x>2$.  Similarly,
the discontinuous part of $f(x')$ results from the production of
domains of length $x$ via aided coalescence (process D above) which
can only occur for $x<2$.  These two effects are of comparable (not
identical) magnitude but opposite sign resulting in only a small
discontinuity in $\dP^{(1)}(x)$. In contrast, the discontinuity in
$\cP^{(1)}(x)$ arises from a {\rm single} process, the origin of which
can be seen from equation~\ref{eqn:series_scaling}.  The second term
on the RHS describes the production of domains of length $x$ via
coalescence of an active domain of length $x'$ with a right neighbor
of length $x-x'$.  However if $x-x'$ is itself active, the right
neighbor can relax before the $x'$ domain, resulting in a domain of
length $>x$.  The rate of producing domains of length $x$ is thus
reduced, as encoded in third term on the RHS. The cutoff in the active
distribution at $x=2$ means that the derivative of this term is
discontinuous at $x=4$. When transferred to the next coarsening stage
(encoded in the LHS of the equation), this discontinuity appears in
$\cP^{(1)}(2)$.  The higher convolutions not shown in
equation~\ref{eqn:series_scaling} are smooth enough not to affect this
argument.

We finally return to justify our two assumptions: firstly, that the
population of $-1$ spins is negligible in the steady state. As noted
above, we have already numerically observed this population to be
$O(\eps)$ (small); we are now in a position to show this
theoretically, as follows. Over a time-scale $\lae 1/\gdot$, the only
mechanism in which $-1$ spins can be created involves propagation of
the facilitating state ($-1$ or 1), via constrained thermal
activation, to within a distance $d<d_c$ to the right of one of the
existing facilitating spins.  Denoting by $\tau_{\rm c}(d)$ the
time-scale on which domains of length $d$ coarsen, the time-scale upon
which a $-1$ spin is created at a distance $d$ from an existing
facilitator is $\tau_{\rm c}(d)/\epsilon$. Once created, such a spin
will relax back to the 0 state on a time-scale $\tau_{\rm c}(d)$.
Hence in the steady state, the population of $-1$ states must be
$O(\epsilon)$, which is indeed small in the limit considered here.

Our second assumption was that of ``independent intervals''. This was
used in~\cite{SolEva99} for the coarsening process of the undriven
chain, and in that case is provably exact.  We have not been able to
prove its strict validity for this driven case.  However, our
simulation results show that the relevant correlation function
$C=\sum_{j=0}^n(d_j-\lav d\rav)(d_{j+1}-\lav d \rav)/\sum_{j=0}^n
(d_j-\lav d\rav)^2$ (in which $d_j$ is the length of $j^{\rm th}$
domain from the left hand end of the chain, and $n$ is the total
number of domains) is not greater than $10^{-3}$ in any of the steady
states considered; the assumption therefore is likely to be
reasonable. Note that our numerical observation of a non-zero
correlation function could still be consistent with the independent
interval approximation being exact in the limit $\eps\to 0$ at fixed
$\nu$.  Indeed, we have observed that, as $\eps$ is tracked downwards,
the numerical value of the correlation function gets smaller.

\section{Summary and Outlook}
\label{sec:summary}

We have incorporated driving dynamics into the kinetically constrained
spin chain of\cite{JaecEis91}, and presented simulation results
showing that the coarsening dynamics of the undriven chain (as solved
in\cite{SolEva99}) are interrupted by steady driving. Consistent with
the broader glassy
literature~\cite{CugKurLeDouPel97,HerGriSol86,CriSom87,Horner96,Thalmann98,BerBarKur00,long_El,SolLeqHebCat97}
we have found that steady driving stabilises the chain at an apparent
age set by the inverse driving rate. We have presented a theory for
this steady state, demonstrating it to be in excellent agreement with
our simulation results. We have shown that a scaling state is
approached at small $\gdot$, and that this scaling state has very
similar properties to its counterpart scaling state reached at long
times in the coarsening chain.

We now outlook some possible directions for future work. Above, we
focused on a chain which is steadily driven at a constant rate
$\gdot$. We also noted the loose analogy of this driving scenario to
that of constant shear rate in a rheological system. In the spirit of
this rheological connection, we can identify possible analogues of two
other standard rheological tests -- step strain and step stress --
which it would be interesting to investigate further. For a step
strain of size $\gamma_0$ we promote, at the time $\tw$ of strain
application, a fraction $\gamma_0$ of $-1$ spins (chosen randomly)
according to $s_i:-1\to 0$ and of $0$ spins (again chosen randomly)
$s_i:0\to 1$.  We do this {\em without regard to the kinetic
  constraint}.  For all other times the system merely relaxes under
its undriven constrained dynamics.  (Note that the step at $\tw$ is
just the ``impulsive limit'' of the above steady shear case: $\gdot
dt=\gamma_0$, with $dt\to 0$ and $\gdot\to\infty$.)  For a step stress
of size $\sigma_0$, we apply the same dynamics just defined for the
step strain up until the time $\tw^+$.  (We can merely rename
$\gamma_0$ by $\sigma_0$ because the ``spring constant'' $k\equiv 1$.)
For $t>\tw$ we implement the ``constant'' strain-rate dynamics defined
above, but with $\gdot$ continuously adjusted to ensure that
$\sigma_0$ remains (on average) a constant.

As noted in the introduction, it would also be interesting to study
FDT in the driven steady state (at constant $\gdot$) to see if any
effective temperature emerges, and (if so) whether it coincides with
any effective FDT temperature of  a coarsening chain of age
$\tw=1/\gdot$.  This is the subject of a forthcoming
publication\cite{upcoming_fdt}.

{\bf Acknowledgements}: The author thanks M.\ E.\ Cates, M.\ R.\ Evans, P.\ D.\ Olmsted and P.\ Sollich for helpful discussions, and EPSRC for financial
support.


\begin{thebibliography}{10}

\bibitem{present_address}
Present address: Department of Physics and Astronomy $\&$ Polymer IRC,
  University of Leeds, Leeds, LS2 9JT, United Kingdom.

\bibitem{upcoming_fdt}
S M Fielding. In preparation.

\bibitem{Barrat98}
A~Barrat.
\newblock {Monte} {Carlo} simulations of the violation of the
  fluctuation-dissipation theorem in domain growth processes.
\newblock {\em Phys.\ Rev.\ E}, 57(3):3629--3632, 1998.

\bibitem{BarBer01}
J~L Barrat and L~Berthier.
\newblock Fluctuation-dissipation relation in a sheared fluid - art. no.
  012503.
\newblock {\em Phys.\ Rev.\ E}, 6301(1):2503--+, 2001.

\bibitem{BerBarKur00}
L~Berthier, J~L Barrat, and J~Kurchan.
\newblock A two-time-scale, two-temperature scenario for nonlinear rheology.
\newblock {\em Phys.\ Rev.\ E}, 61(5):5464--5472, 2000.

\bibitem{BouCugKurMez98}
J~P Bouchaud, L~F Cugliandolo, J~Kurchan, and M~M{\'{e}}zard.
\newblock Out of equilibrium dynamics in spin-glasses and other glassy systems.
\newblock In A~P Young, editor, {\em Spin glasses and random fields},
  Singapore, 1998. World Scientific.

\bibitem{BraDerGod94}
A~J Bray, B~Derrida, and C~Godreche.
\newblock Nontrivial algebraic decay in a soluble model of coarsening.
\newblock {\em Europhys.\ Lett.}, 27(3):175--180, 1994.

\bibitem{CipManBalWei00}
L~Cipelletti, S~Manley, R~C Ball, and D~A Weitz.
\newblock Universal aging features in the restructuring of fractal colloidal
  gels.
\newblock {\em Phys.\ Rev.\ Lett.}, 84(10):2275--2278, 2000.

\bibitem{CloBorLei00}
M~Cloitre, R~Borrega, and L~Leibler.
\newblock Rheological aging and rejuvenation in microgel pastes.
\newblock {\em Phys.\ Rev.\ Lett.}, 85(22):4819--4822, 2000.

\bibitem{CriRitRocSel00}
A~Crisanti, F~Ritort, A~Rocco, and M~Sellitto.
\newblock Inherent structures and nonequilibrium dynamics of one- dimensional
  constrained kinetic models: a comparison study.
\newblock {\em J.\ Chem.\ Phys.}, 113(23):10615--10634, 2000.

\bibitem{CriSom87}
A~Crisanti and H~Sompolinsky.
\newblock {\em Phys.\ Rev.\ A}, 36:4922, 1987.

\bibitem{CugKur93}
L~F Cugliandolo and J~Kurchan.
\newblock Analytical solution of the off-equilibrium dynamics of a long- range
  spin-glass model.
\newblock {\em Phys.\ Rev.\ Lett.}, 71(1):173--176, 1993.

\bibitem{CugKur94}
L~F Cugliandolo and J~Kurchan.
\newblock On the out-of-equilibrium relaxation of the
  {Sherrington}-{Kirkpatrick} model.
\newblock {\em J.\ Phys.\ A}, 27(17):5749--5772, 1994.

\bibitem{CugKurLeDouPel97}
L~F Cugliandolo, J~Kurchan, P~Le{D}oussal, and L~Peliti.
\newblock Glassy behaviour in disordered systems with nonrelaxational dynamics.
\newblock {\em Physical Review Letters}, 78(2):350--353, 1997.

\bibitem{CugKurPel97}
L~F Cugliandolo, J~Kurchan, and L~Peliti.
\newblock Energy flow, partial equilibration, and effective temperatures in
  systems with slow dynamics.
\newblock {\em Phys.\ Rev.\ E}, 55(4):3898--3914, 1997.

\bibitem{DerAjdDucLeq00}
C~Derec, A~Ajdari, G~Ducouret, and F~Lequeux.
\newblock Rheological characterization of aging in a concentrated colloidal
  suspension.
\newblock {\it Compte Rendus d'Acad. des Sciences (Paris)}. Accepted for
  publication, 2000.

\bibitem{DerAjdLeq99}
C~Derec, A~Ajdari, and F~Lequeux.
\newblock Mechanics near a jamming transition: a minimalist model.
\newblock {\em Faraday Discuss.}, (112):195--207, 1999.

\bibitem{HebLeq98}
P~H{\'{e}}braud and F~Lequeux.
\newblock Mode-coupling theory for the pasty rheology of soft glassy materials.
\newblock {\em Phys.\ Rev.\ Lett.}, 81(14):2934--2937, 1998.

\bibitem{HerGriSol86}
J~A Hertz, Grinstein G, and S~Solla.
\newblock In J~L van Hemmen and I~Morgenstern, editors, {\em Proceedings of the
  Heidelberg Colloquium on Glassy Dynamics and Optimization}, Berlin, 1987.
  Springer Verlag.

\bibitem{Horner96}
H~Horner.
\newblock Drift, creep and pinning of a particle in a correlated random
  potential.
\newblock {\em Zeitschr.\ Phys.\ B}, 100(2):243--257, 1996.

\bibitem{JaecEis91}
J~J{\"{a}}ckle and S~Eisinger.
\newblock A hierarchically constrained kinetic {Ising}-model.
\newblock {\em Zeitschr.\ Phys.\ B}, 84(1):115--124, 1991.

\bibitem{KnaBelMunViaLeqHar00}
A~Knaebel, M~Bellour, J~P Munch, V~Viasnoff, F~Lequeux, and J~L Harden.
\newblock Aging behavior of laponite clay particle suspensions.
\newblock {\em Europhys.\ Lett.}, 52(1):73--79, 2000.

\bibitem{KobBar00}
W~Kob and J~L Barrat.
\newblock Fluctuations, response and aging dynamics in a simple glass- forming
  liquid out of equilibrium.
\newblock {\em Eur.\ Phys.\ J.\ B}, 13(2):319--333, 2000.

\bibitem{MarParRicRui98}
E~Marinari, G~Parisi, F~Ricci-Tersenghi, and J~J Ruiz-Lorenzo.
\newblock Violation of the fluctuation-dissipation theorem in finite-
  dimensional spin glasses.
\newblock {\em J.\ Phys.\ A-Math.\ Gen.}, 31(11):2611--2620, 1998.

\bibitem{MonBou96}
C~Monthus and J~P Bouchaud.
\newblock Models of traps and glass phenomenology.
\newblock {\em J.\ Phys.\ A}, 29(14):3847--3869, 1996.

\bibitem{RamCip01}
L~Ramos and L~Cipelletti.
\newblock Ultraslow dynamics and stress relaxation in the aging of a soft
  glassy system - art. no. 245503.
\newblock {\em Phys.\ Rev.\ Lett.}, 8724(24):5503--+, 2001.

\bibitem{Reichl80}
L~E Reichl.
\newblock {\em A modern course in statistical physics}.
\newblock University of Texas Press, Austin, 1980.

\bibitem{RicParStaAre01}
F~Ricci-Tersenghi, G~Parisi, D~A Stariolo, and J~J Arenzon.
\newblock Comment on ''two time scales and violation of the fluctuation-
  dissipation theorem in a finite dimensional model for structural glasses'' -
  reply.
\newblock {\em Phys.\ Rev.\ Lett.}, 86(20):4717--4717, 2001.

\bibitem{long_El}
P~Sollich.
\newblock Rheological constitutive equation for a model of soft glassy
  materials.
\newblock {\em Phys.\ Rev.\ E}, 58:738--759, 1998.

\bibitem{SolEva99}
P~Sollich and M~R Evans.
\newblock Glassy time-scale divergence and anomalous coarsening in a
  kinetically constrained spin chain.
\newblock {\em Phys.\ Rev.\ Lett.}, 83(16):3238--3241, 1999.

\bibitem{SolLeqHebCat97}
P~Sollich, F~Lequeux, P~H{\'{e}}braud, and M~E Cates.
\newblock Rheology of soft glassy materials.
\newblock {\em Phys.\ Rev.\ Lett.}, 78:2020--2023, 1997.

\bibitem{Thalmann98}
F~Thalmann.
\newblock {\em Eur.\ Phys.\ J.\ B}, 3:497, 1998.

\end{thebibliography}

\end{document}